\documentclass[12pt]{iopart}
\usepackage{iopams}  
\usepackage{setstack}  
\usepackage[utf8]{inputenc}
\usepackage{graphicx}
\usepackage{xcolor}
\usepackage{ulem}

\begin{document}

\newcommand{\edits}{\color{red}} 
\newcommand\editssout{\bgroup\markoverwith{\textcolor{red}{\rule[0.5ex]{2pt}{1.5pt}}}\ULon} 

\newcommand{\LC}{\color{green}}  
\newcommand{\TR}{\color{blue}}  
\newcommand{\TRcut}{\color{blue}}  
\newcommand\TRsout{\bgroup\markoverwith{\textcolor{blue}{\rule[0.5ex]{2pt}{1.5pt}}}\ULon}

\renewcommand\thefigure{\arabic{figure}}    

\title[An equilibrium model for ribosome competition]{An equilibrium model for ribosome competition}

\author{Pascal S. Rogalla}
\address{Institute for biological and medical engineering, Schools of
engineering, biology and medicine, Universidad Catolica de Chile,
Chile}
\address{Department of chemical and bioprocess engineering, School of
engineering, Universidad Catolica de Chile, Chile}
\address{I. Physikalisches Institut (IA), RWTH Aachen University, 52074 Aachen, Germany}
\ead{pascal.rogalla@rwth-aachen.de}

\author{Timothy J. Rudge}
\address{Institute for biological and medical engineering, Schools of
engineering, biology and medicine, Universidad Catolica de Chile,
Chile}
\address{Department of chemical and bioprocess engineering, School of
engineering, Universidad Catolica de Chile, Chile}

\author{Luca Ciandrini}
\address{CBS, Universit\'e de Montpellier, CNRS and INSERM, Montpellier, France}
\address{Laboratoire Charles Coulomb (L2C), Universit\'e de Montpellier and CNRS, Montpellier, France}
\ead{luca.ciandrini@umontpellier.fr}

\begin{abstract}
The number of ribosomes in a cell is considered as limiting, and gene expression is thus largely determined by their cellular concentration. In this work we develop a toy model to study the trade-off between the ribosomal supply and the demand of the translation machinery, dictated by the composition of the transcript pool. Our equilibrium framework is useful to highlight qualitative behaviours and new means of gene expression regulation determined by the fine balance of this trade-off. We also speculate on the possible impact of these mechanisms on cellular physiology.
\end{abstract}

\vspace{2pc}
\maketitle

\section{Introduction}
Gene expression is a costly process which requires the constant involvement of the components of its transcriptional and translational machineries (polymerases, transcription factors, ribosomes, tRNAs,...) for protein synthesis. The interplay between cellular resources and protein synthesis is a topic that has recently drawn the attention of a growing community (see for instance~\cite{Brewster2014a, deJong2017ResourceMachinery, Ceroni2018Burden-drivenExpression, Darlington2018DynamicGenes, Sabi2019ModellingExpression, Lin2018HomeostasisCells, Friedlander2016IntrinsicCrosstalk, Qian2017ResourceCircuits, Das2017EffectNoise, Landman2017Self-consistentArchitectures} to mention just a few recent publications), and its understanding is pivotal to the development of a quantitative characterisation of gene expression regulation. This could also unlock applications related to, for instance, the design of gene circuits, the control of the cellular physiology or optimal protein synthesis for biotechnological applications. \\
When modelling transcription and translation, cellular resources such as transcription factors and RNA polymerases are often considered as being unlimited, with just a few recent exceptions~\cite{Brewster2014a, Rydenfelt2014StatisticalTitration, Weinert2014ScalingFugacity}. At the translational level, ribosomes are known to be limiting~\cite{Shah2013Rate-limitingTranslation} and their abundance is strongly related to expression levels and cellular physiology~\cite{Scott2010, Scott2011, Scott2014}. However, even state-of-the-art tools such as the Ribosome Binding Site (RBS) calculator~\cite{Salis2009}, which is able to predict the mRNA's recruitment rate of a ribosome given the nucleotide sequence of the transcript, neglect the possible depletion effects that strong synthetic RBSs might have on the ribosomal pool, and thus potentially induce protein burden.\\ 
For all these reasons here we develop a toy model to study the trade-off between demand (mRNA abundances) and supply of resources (ribosomes). Although neglecting many other important aspects of resource competition and cell physiology -for instance \hbox{metabolism-,} we show that even this simple coarse grained model predicts a rich behaviour and new means of gene expression regulation just by considering ribosome trade-off between different mRNA populations. Furthermore, our model provides an intuitive mechanistic explanation of the phenomenological growth law relating ribosome abundances and growth rates.

In the next section we first revise an equilibrium model of gene expression that we take as a reference point for our analysis, then generalise it to adapt it to the translation process in which several ribosomes can concurrently translate an mRNA. In the last part of the paper we speculate on how the model predictions can be interpreted to rationalise gene expression regulation by the trade-off between supply and demand of ribosomal resources. \\

\section{Model}
\subsection{Thermodynamic model of gene expression}
\label{subs:thermodynamic}
In this section we summarise the basics of a thermodynamic model of gene expression based on the assumption that transcription occurs when an RNA polymerase (RNAP) is bound to a promoter of interest~\cite{Bintu2005, Bintu2005a}. Equilibrium statistical mechanics provides the tools to compute the probability $p$ that an RNAP is bound to a promoter, and gene expression levels are assumed to be proportional to that quantity. In more detail, if $m$ is the concentration of mRNAs and $k_\mathrm{TX}$ their transcription rate, one can write 
\begin{equation}
    \frac{d m}{dt} = k_\mathrm{TX} \, p - \gamma m \,,
    \label{eq:transcription}
\end{equation}
where $\gamma$ is the degradation rate of mRNAs. At the steady-state one then finds that $m =  p\, k_\mathrm{TX} / \gamma$, which can be used as a proxy for gene expression.

Interactions with {\it cis-} or {\it trans-} regulatory elements will not be considered here since they are not related to the topic of our work, and detailed reviews can be found in~\cite{Phillips2015NapoleonEquilibrium, Phillips2012PhysicalCell}. Instead, here we focus only on the role of particle (RNAP, ribosome) competition, and show how relative expression between different genes could arise purely by trade-off between those elements.\\

To introduce the model we follow the procedure established in~\cite{Bintu2005} that considers a single {\it specific site} (playing the role of a promoter and characterised by an occupation number $n_b$) immersed in a pool of $n$ particles (RNAPs) and $N$ {\it non-specific sites} (DNA sites other than the promoter of interest). The specific site can be occupied by a particle ($n_b = 1$) or be empty ($n_b = 0$), and the remaining $n_f = n - n_b$ particles are distributed in the $N$ non-specific sites. 
Since the expression rate of the gene of interest is directly related to the occupation of the specific site, we need to compute the probability $p(n_b = 1)$ of finding the particle bound specifically. 
For the sake of completeness, below we re-derive the results of this model, but we refer for instance to~\cite{Bintu2005, Phillips2015NapoleonEquilibrium, Phillips2012PhysicalCell} for more detailed explanations.
\\
According to equilibrium statistical mechanics, the probability of the macrostate $j$ with free energy $E(j)$ is given by the Boltzmann distribution $p(j)=D(j)e^{-\beta E(j)}/Z^{tot}$, with $\beta=1/k_BT$ being the inverse temperature and $D(j)$ the degeneracy of the macrostate -the number of microstates with the same energy. The sum of all partition functions of the macrostates is the total partition function $Z^{tot}$, which gives the normalisation factor of the probability so that $\sum_j p(j) = 1$.
By identifying the macrostate with the occupation number $n_b$ we are then able to compute $p(n_b=1)$ knowing $\epsilon_b := E(n_b=1)$ and $\epsilon_f := E(n_b=0)$.

In order to simplify the mathematical expressions, the partition functions $Z(j)$ and the degeneracies $D(j)$ are usually written in terms of a reference state, here $n_b=0$, and indicated with the subscript $0$: $Z_0(n_b) := Z(n_b)/Z(n_b=0)$ and $D_0(n_b) := D(n_b)/D(n_b=0)$. The expression for $p(n_b=1)$ can therefore be written as
	\begin{equation}
	    p(n_b=1) =\frac{Z_0(n_b=1)}{Z_0^\mathrm{tot}}=\frac{D_0(n_b=1)e^{-\beta \Delta\epsilon}}{1+D_0(n_b=1)e^{-\beta \Delta\epsilon}}
	    \label{Sec2.1_Eq1}\,\,,
	\end{equation}
where 
$\Delta\epsilon := \epsilon_b-\epsilon_f$ stands for the energy turnover by the process of binding the particle. \\
The degeneracy $D(j)$ counts the possible ways to arrange the non-specifically bound RNAP to the non-specific sites in the case of state $j$. Here, the non-specific sites are assumed to be of the order of the number of base pairs not belonging to the gene of interest (specific-site). Hence $D_0(n_b=1)\simeq n/N$ (assuming $N \gg n$), that can be plugged in Eq.~(\ref{Sec2.1_Eq1}) to obtain 
	\begin{equation}
	    p(n_b=1) =\frac{\frac{n}{N}e^{-\beta \Delta\epsilon}}{1+\frac{n}{N}e^{-\beta \Delta\epsilon}}
	    \label{eq:p_thermo}\,\,.
	\end{equation}

\subsection{mRNA translation and notations}

Our model is a direct extension of this equilibrium framework. Since we present it in terms of mRNA translation, we will consider ribosomes instead of RNAPs, landing on ribosome binding sites or 5'UTRs (untranslated regions) instead of promoters. However, we keep the same notations because of the generality of the mathematical structure underlying the results, that could then immediately re-interpreted in other systems (as well as transcription).\\

The mRNA pool is considered to be in equilibrium with a reservoir of $n_f$ free ribosomes diffusing in the cytoplasm. The volume available to the ribosomes in the reservoir then can be regarded as the ensemble of non-specific sites. Defining a 3D-grid, each site being approximately of the volume of a ribosome,  provides the statistical means for a detailed description of the degeneracy $D(j)$. Similar  ideas for, i.e., ligand-receptor binding on a 2D-grid have motivated this approach~\cite{Phillips2012PhysicalCell}.
A microstate is a realisation of the distribution of the diffusing or free ribosomes $n_f$ into the $N$ compartments, Fig.~\ref{fig:sketch}(a). The number of possible microstates for such a system is simply given by the binomial coefficient $D(n_b)={N\choose n-n_b}$. \\

In the most general case we will consider $\mathcal{N}$ populations of mRNAs, each one characterised by a different ribosome affinity, $\Delta\epsilon^{(i)}$, and competing for the $n_f$ ribosomes in the reservoir. The mRNA population $i$ contains $M^{(i)}$ transcripts, with a maximal capacity $n_\mathrm{max}^{(i)}$ corresponding to largest number of translating ribosomes that an mRNA can fit. Hence, $n_\mathrm{max}^{(i)} := L/\ell$ with $L$ being the length of the transcript in codons, and $\ell$ the footprint of the ribosomes. A typical gene has a length $L=300$ codons and a ribosome covers $\ell=10$ codons.
We summarise in Table~\ref{table:notations} the symbols used in this work.

\begin{table}[h!]
\centering
\begin{tabular}{l|p{8cm}}
 \textbf{Symbol} & \textbf{Meaning}  \\ \hline\hline
 $n_f$          &   Unbound, free ribosomes constituting the reservoir \\
 $n_b^{(i)}$    &   Ribosomes bound to all mRNAs belonging to the population $i$ \\
 $n_\mathrm{max}^{(i)}$     &   Ribosome capacity of an individual mRNA belonging to the population $i$\\
 $N$    &    Number of cytoplasmic compartments (non specific sites)\\
 $M^{(i)}$      &   Number of mRNAs in population $i$\\
 $\mathcal{N}$  &   Number of mRNA populations\\
 $n_b$          &   Ribosomes bound (total): $ n_b= \sum_{i=0}^{\mathcal{N}}n_b^{(i)}$ \\
 $n$            &   Total number of ribosomes: $n=n_f + n_b$ \\
 $\Delta\epsilon^{(i)}$ & Energy turnover when a single ribosome binds to an mRNA of population $i$ \\

 \hline
\end{tabular}
\caption{Summary of the symbols used and their meaning.}
\label{table:notations}
\end{table}

\subsection{Extension to many particle binding}
\label{subs:cases}
In the previous sections we revised the equilibrium model of gene expression as introduced in~\cite{Bintu2005, Bintu2005a} in the context of transcription, and which has been extended in~\cite{Brewster2014a, Rydenfelt2014StatisticalTitration, Weinert2014ScalingFugacity} to multiple genes to investigate the titration of transcription factors. Here we extend this model to consider the concurrent binding of many particles on the same substrate. This is a natural extension when studying the competition of genes for RNAP -in the case of transcription- and for ribosomes - in the case of translation. As also explained above, we will develop this model in the context of mRNA translation, where ribosomes are considered as limiting~\cite{Shah2013Rate-limitingTranslation} and their abundance strictly related to the cellular physiology~\cite{Scott2010, Scott2011, Scott2014}. 

In order to do that, instead of considering that each mRNA is either occupied by one ribosome ($n_b=1$, actively translating state) or empty ($n_b=0$, untranslating state) as it would be in the standard thermodynamic model introduced in Section~\ref{subs:thermodynamic}, we imagine that a number $n_b \leq n_\mathrm{max}$ of ribosomes can be recruited by an individual mRNA, as shown in Figure~\ref{fig:sketch}(a). This corresponds to assume that the RBS of the mRNA becomes immediately available (if $n_b < n_\mathrm{max}$) and a new ribosome can bind again the lattice. 

\begin{figure}[hbt]
    \centering
    \includegraphics[width=.9\linewidth]{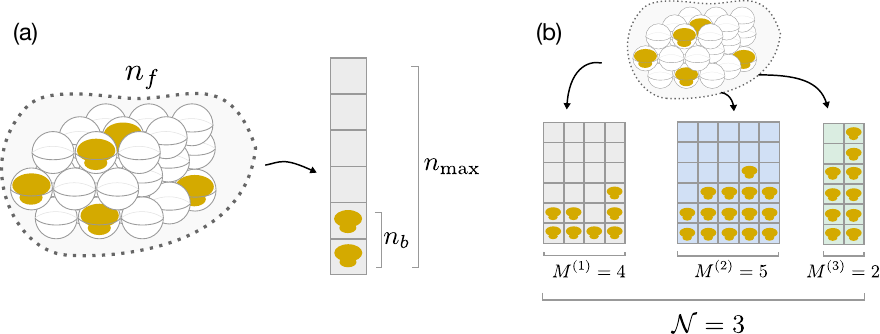}
    \caption{(a) Sketch of a possible microstate of the model, with $n_f$ free ribosomes (yellow particles) in the $N$ cytoplasmic compartments (the white spheres) and $n_b = 2$ ribosomes on the mRNA (grey squares). With respect to the $n_b = 0$ situation, this configuration has an energy turnover $n_b\Delta\epsilon$. $n_\mathrm{max}$ indicates the maximum number of ribosomes concurrently translating the mRNA (number of squares). (b) The full model takes into account a number $\mathcal{N}$ of mRNA populations (here $\mathcal{N}=3$, each one composed of $M^{(i)}$ mRNAs with the same $n_{max}$).\label{fig:sketch}}
    
\end{figure}

We point out that, due to the model definitions, the recruitment capability of an mRNA decreases with its ribosome occupation as its number of available sites decreases -see~\ref{appendix:Gillespie}. This is somehow consistent with what other models considering ribosome interference would predict~\cite{greulich_mixed_2012}.

In the following sections we will develop the model with multiple ribosome binding on an individual mRNA (Section~\ref{subsubs:cases_singlecopy}), then extend it to several copies of the same transcript (Section~\ref{subsubs:cases_onepopulation}) and finally extend it to the case of several mRNA types having different properties. This general case depicted in Figure~\ref{fig:sketch}(b) will be treated in Section~\ref{subsubs:cases_generalcase}.

\subsubsection{A single mRNA type, one copy.} \label{subsubs:cases_singlecopy}
The model revisited in Section \ref{subs:thermodynamic} illustrates the derivation of the occupation probability $p$ for the simple case of two states: the single specific site being occupied or unoccupied. This case corresponds to $n_\mathrm{max}=1$, and here we extend it to any value of $n_\mathrm{max}$ in order to obtain the occupation probability $p(n_b)$ for a system with $n_\mathrm{max}+1$ possible states. In this section we begin considering an individual mRNA ($\mathcal{N}=1$, $M=1$) in contact with a reservoir of ribosomes.\\
In the case of the equilibrium thermodynamic model (Section~\ref{subs:thermodynamic}), gene expression levels are proportional to $p(n_b=1)$, as at most one particle can bind the promoter. Instead, in our model extension with multiple particle binding we can reasonably assume that protein synthesis levels are proportional to $\big<n_b\big>$, the average number of ribosomes bound to the transcript. This is what it is usually done to interpret experimental data in mRNA translation, and a measure of translation efficiency is directly related to ribosome density~\cite{Li2015HowEfficiency}. 

The partition functions $Z_0(n_b) = D_0(n_b) \, e^{-\beta [n_b\Delta\epsilon]}$ for each macrostate characterised by $n_b$ are necessary to compute $\big< n_b \big>$ and are derived in \ref{appendix:DerivationGeneralCase} 
for the most general case. The reader can recover the results of this current section by fixing $\mathcal{N}=1$, $M=1$ in the equations of this appendix. As multiple binding can in principle be considered with a capacity $n_\mathrm{max}$ larger than the total number of particles $n$, we further need to consider that the mRNA cannot be occupied by more ribosomes than there are available for binding. Summing up the different accessible states and by taking into account this restriction we obtain
\begin{equation}
Z_0^\mathrm{tot} = \sum_{n_b=0}^{\min(n,n_\mathrm{max})}Z_0(n_b)\,\,.
\label{Case1_Eq3}
\end{equation}
%
For a macrostate with partition function $Z_0(n_b)$ we find two separable parts contributing to the degeneracy $D_0(n_b)$. On the one hand, the $n_f$ free ribosomes are distributed into the $N$ cytoplasmic compartments. On the other hand, there is an additional distribution by the $n_b$ bound ribosomes on the mRNAs' lattice. As the lattice can be maximally occupied by a number $n_\mathrm{max}$ of particles, the possible ways of distributing $n_b$ bound ribosomes is given by the binomial coefficient ${n_\mathrm{max} \choose n_b}$. The degeneracy of the macrostate then is $D(n_b)= {N \choose n - n_b} {n_\mathrm{max} \choose n_b}$ and the normalised degeneracy for the single mRNA case depicted in Figure~\ref{fig:sketch}(a) reads
\begin{eqnarray}
D_0(n_b) &:=
\left\{
\begin{array}{ll}
\frac{{N \choose n - n_b} {n_\mathrm{max} \choose n_b}}{{N \choose n}} = \prod_{j=1}^{n_b}\left[\frac{(n-j+1)}{(N-n+j)}\frac{(n_\mathrm{max}-j+1)}{j}\right]& n_b >0 \\
1 & n_b=0\,\,.\label{Case1_Eq2}
\end{array}
\right. 
\end{eqnarray}
We also recall that $D(n_b)$ is strictly related to the hypergeometric distribution, as the problem studied in this section is mathematically equivalent to draw $n$ marbles (bound ribosomes) from an urn containing $N+n_\mathrm{max}$ marbles of which $N$ are white (cytoplasmic compartments) and $n_\mathrm{max}$ are black (specific ribosome binding site). We again refer to~\ref{appendix:DerivationGeneralCase} for details.\\

The quantity $\big<n_b\big>$, proportional to the translation efficiency, can hence be computed as the expected value of all $n_\mathrm{max}+1$ possible states of energy with probabilities $p(n_b)$:
\begin{equation}
\big< n_b \big> = \sum_{n_b=0}^{\min(n,n_\mathrm{max})}n_b\,p(n_b)=\sum_{n_b=0}^{\min(n,n_\mathrm{max})}n_b\,\frac{Z_0(n_b)}{Z_0^\mathrm{tot}}\,\,.
\label{Case1_Eq4}
\end{equation}
As an example, Figure \ref{Case1_Fig1} shows the behaviour of $\big<n_b\big>$ as a function of the total number of ribosomes $n$ for three exemplifying values $\beta\Delta\epsilon$. The $\Delta\epsilon$ represents the strength of the RBS and accounts for hybridization and non-optimal spacing of ribosome subunits, unfolding of mRNA, etc.~\cite{Salis2009}. 
\begin{figure}[h!bt]
	\centering
	\includegraphics[width=1\textwidth]{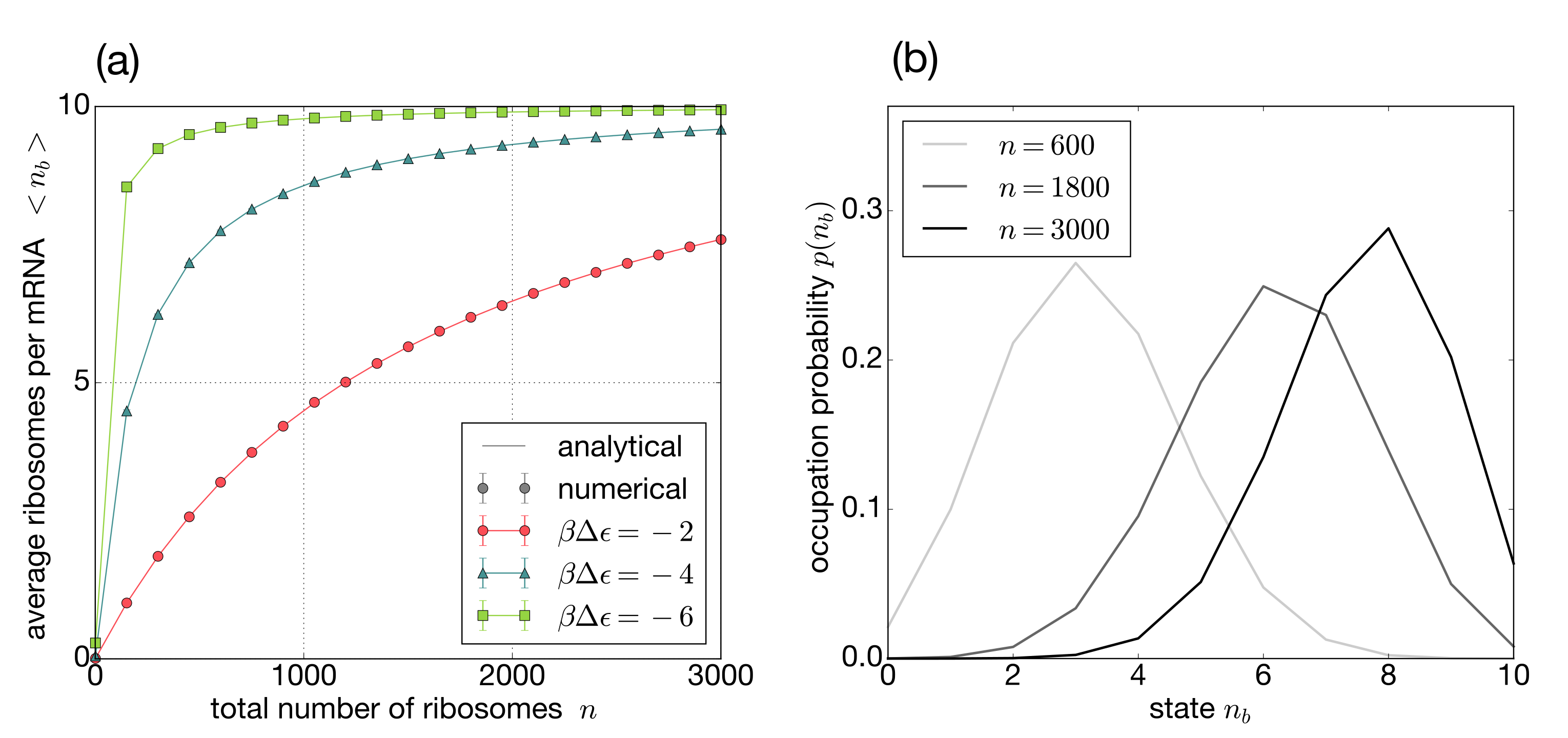}
	\caption{(a)A single mRNA type ($\mathcal{N}=1$), one copy ($M=1$): Plot of $\big<n_b\big>$ as a function of $n\in(0,3000]$ for three different values $\beta\Delta\epsilon \in\{-2, -4, -6\}$. The number of cytoplasmic compartments is fixed to $N=10^4$ and the maximal ribosome capacity per mRNA is $n_\mathrm{max}=10$. The solid line is obtained with the analytic solution Eq.~(\ref{Case1_Eq4}), whereas the circles with error-bars are obtained by simulating the system using the Gillespie chemical reaction algorithm~\cite{Gillespie1977ExactReactions}. (b) occupation probability $p(n_b)$ for $\beta\Delta\epsilon=-2$.}
	\label{Case1_Fig1}
\end{figure}
The analytical solution Eq.~(\ref{Case1_Eq4}) is compared to simulations of the system with the Gillespie algorithm, whose details are explained in~\ref{appendix:Gillespie}. The agreement between the theory and the numerical simulation is excellent.

\subsubsection{A single mRNA population, $M$ copies.}
\label{subsubs:cases_onepopulation}
For the sake of completeness we go one step further, and extend the previous case of a single mRNA copy $M=1$ to any $M \geq 1$. Substantially, one obtains the same Equations (\ref{Case1_Eq3}), (\ref{Case1_Eq2}) and (\ref{Case1_Eq4}) with the substitution $n_{max} \to n_{max}\,M$, meaning that the demand of the transcripts for the ribosomes is increased of a factor $M$. We define  $\big<\bar{n_b}\big> := \big<n_b\big>/(n_\mathrm{max}M)$ being the occupation (normalised) of each individual mRNA, and we also refer to $\big<\bar{n_b}\big>$ as the translation efficiency of an mRNA belonging to this population.\\
Figure~\ref{Case2_Fig1}(a) shows $\big<\bar{n_b}\big>$ as a function of the total number of ribosomes $n$ with a number of $M=10$ mRNA copies. In order to show the competition between different members of the same mRNA population, we compare the outcomes with the ones obtained for the single-copy case treated in the previous section and in Figure~\ref{Case1_Fig1}, and the analytical solution for $M=1$ case is shown by the dashed lines. 
\begin{figure}[h!bt]
	\centering
	\includegraphics[width=1\textwidth]{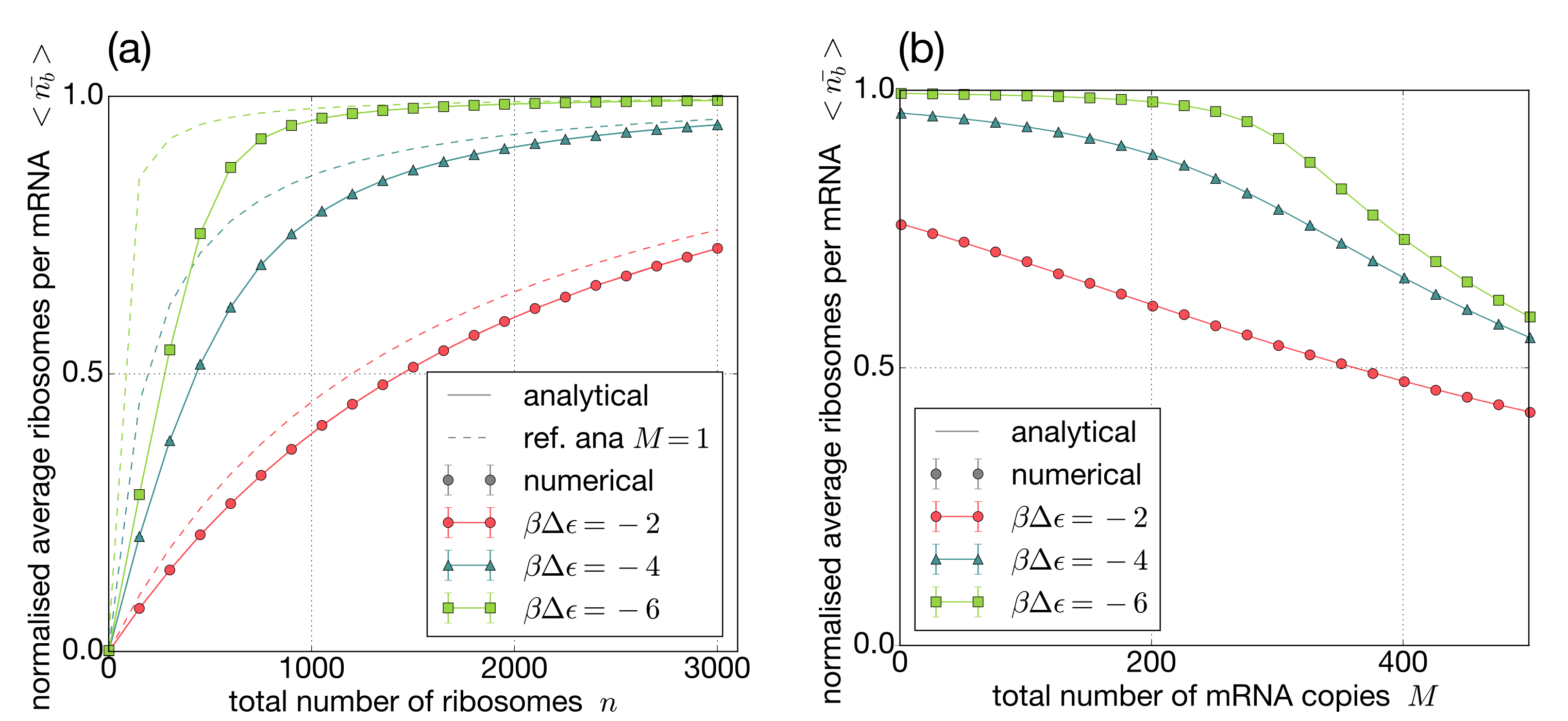}
		\caption{Unique mRNA population $\mathcal{N}=1$, $M\geq1$. Normalised transcript occupancy $\big<\bar{n_b}\big>$ as a function of the total number of ribosomes $n$ (a) for fixed $M=50$, and of the total number of transcripts $M$ (b) for fixed $n=3000$. Dashed lines in panel (a) show the results for $M=1$. In both panels $n_\mathrm{max}=10$.} 
	\label{Case2_Fig1}
\end{figure}

So far we have considered the {\it supply} of ribosomes $n$ as a control parameter of the model. However, when considering multi-copy mRNAs, we can also vary the {\it demand} of the system, i.e. $M$. When the demand is low, each individual mRNA is at its maximum capacity, and it decreases by increasing the number of mRNAs $M$ in the population -see Fig.~\ref{Case2_Fig1}(b).

\subsubsection{$\mathcal{N}$ mRNA populations, $M^{(i)}$ copies.}
\label{subsubs:cases_generalcase}
The genome codes for a large number of different types of mRNAs with unique properties as RBS strength, codon length and translation rate. To account for this diversity, we develop the general case of $i=1,...,\mathcal{N}$ distinct types of mRNA populations. Where necessary, the superscript $(i)$ indicates that a parameter is specific for the population $i$. \\
The occupation probability $p(n_b^{(l)})$ of having $n_b^{(l)}$ ribosomes bound to the mRNA population $l$ is given by the sum of all states with the given $n_b^{(l)}$ divided by $Z_0^{tot}$:
\begin{equation}
p(n_b^{(l)})=\sum_{n_b^{(l)}=const}\frac{Z_0(n_b^{(1)},...,n_b^{(l)},...,n_b^{(\mathcal{N})})}{Z_0^{tot}}\,\,.
\label{Case3_Eq1}
\end{equation}
We recall that each normalised partition function $Z_0$ represents a unique configuration of energy as a function of the bound particles to the different populations $\sum_{i=1}^{\mathcal{N}}n_b^{(i)}\Delta\epsilon^{(i)}$, with
\begin{eqnarray}
Z_0(\{n_b^{(i)}\})&= D_0(\{n_b^{(i)}\}) \, e^{-\beta [\sum_{i=1}^{\mathcal{N}}n_b^{(i)}\Delta\epsilon^{(i)}]}\,\,,& 
\end{eqnarray}
and
\begin{eqnarray}
D_0(\{n_b^{(i)}\}) &=
\left\{
\begin{array}{ll}
\frac{ {N \choose n-n_b} \prod_{i=1}^{\mathcal{N}} {n_\mathrm{max}^{(i)} M^{(i)} \choose n_b^{(i)} }} {{N \choose n}} = 
\prod_{j=1}^{n_b}\left[\frac{(n-j+1)}{(N-n+j)}\right]\,\prod_{i=1}^{\mathcal{N}}D_0^*(i)& n_b >0 \\
1 & n_b=0\,\,,
\end{array}
\right.& \label{Case3_Eq3}\\
D_0^*(i) &=
\left\{
\begin{array}{ll}
\prod_{k=1}^{n_{b}^{(i)}}\left[\frac{n_{max}^{(i)}M^{(i)}-k+1}{k}\right]& n_{b}^{(i)}>0 \\
1 & n_{b}^{(i)}=0 \,,
\end{array}
\right.& \label{Case3_Eq4}
\end{eqnarray} 	
where $n_b=\sum_{i=1}^{\mathcal{N}}n_b^{(i)}$ is the total number of bound ribosomes. \\
The degeneracy $D_0$ in Eq.~(\ref{Case3_Eq3}) consists of two parts: The left product corresponds to the degeneracy of the case of a single mRNA population with one copy number $(\mathcal{N}=1, M=1)$, see Equation (\ref{Case1_Eq2}); the right part is the product over all populations of distributing $n_b^{(i)}$ particles on the $n_{max}^{(i)}M^{(i)}$ sites available for each population $i$, as sketched in Figure \ref{fig:sketch}(b).  Similarly to Equation (\ref{Case1_Eq2}), this problem is similar to randomly drawing $n$ marbles from an urn containing $\sum^\mathcal{N} n_\mathrm{max}^{(i)} M^{(i)}$ marbles of $\mathcal{N}$ different colors. The degeneracy $D_0$ in Eq.~(\ref{Case3_Eq3}) is thus related to the multivariate generalisation of the hypergeometric distribution.
Another way to understand this is to imagine that a number $n_b^{(l)}$ of ribosomes bound to the population $l$ can be realised by the product of all possible distributions $\{n_b^{(i)}\} i\neq l$ of ribosomes bound to other populations, as given by $\prod_{i=1, i\neq l}^{\mathcal{N}}D_0^*(i)$.\\
Following the very same considerations, one finds the total partition function to be
\begin{equation} 
Z_0^{tot}:= \sum_{n_b^{(1)}=0}^{min(n,n_{max}^{(1)}M^{(1)})}\cdot\,...\cdot \sum_{n_b^{(\mathcal{N})}=0}^{min(n-\sum_{i=1}^{\mathcal{N}-1}n_b^{(i)},n_{max}^{\mathcal{(N)}}M^{(\mathcal{N})})}Z_0(\{n_b^{(i)}\}) \,\,.
\end{equation}

Thanks to the general case developed in this section we can study the competition between populations under limited resources, as their different ribosome recruitment capability can lead to counter-intuitive results. To facilitate the observation of such complex behaviour we compare normalised values of the average protein synthesis per mRNA (the translation efficiency) and per mRNA population relative to all the other populations. 
The effects of competition between different populations can hence be computed with the relative protein synthesis levels
\begin{equation}
\big<n_b^{(l)}\big>_\mathrm{rel} = \frac{\big<{n_b}^{(l)}\big>}{\sum_{i}\big<{n_b}^{(i)}\big>}\,\,.
\label{eq:rel_nb}
\end{equation}

The result of a competition between $\mathcal{N}=2$ populations is shown in Fig.~\ref{Case3_Fig1}. In panel (a) we plot the efficiency $\big<\bar{n_b}\big>$ of the two populations as a function of the total number of ribosomes $n$. One can notice that, as expected, the population with the highest affinity for ribosomes reaches its saturation faster compared to the other transcripts, and only then the second population increases their translation efficiency. This can also be appreciated by observing the relative total amount of ribosomes $\big<n_b^{(i)}\big>_\mathrm{rel}$ involved in translation of the two populations, panel (b). 

	\begin{figure}[hbt] 
		\centering
		\includegraphics[width=1\textwidth]{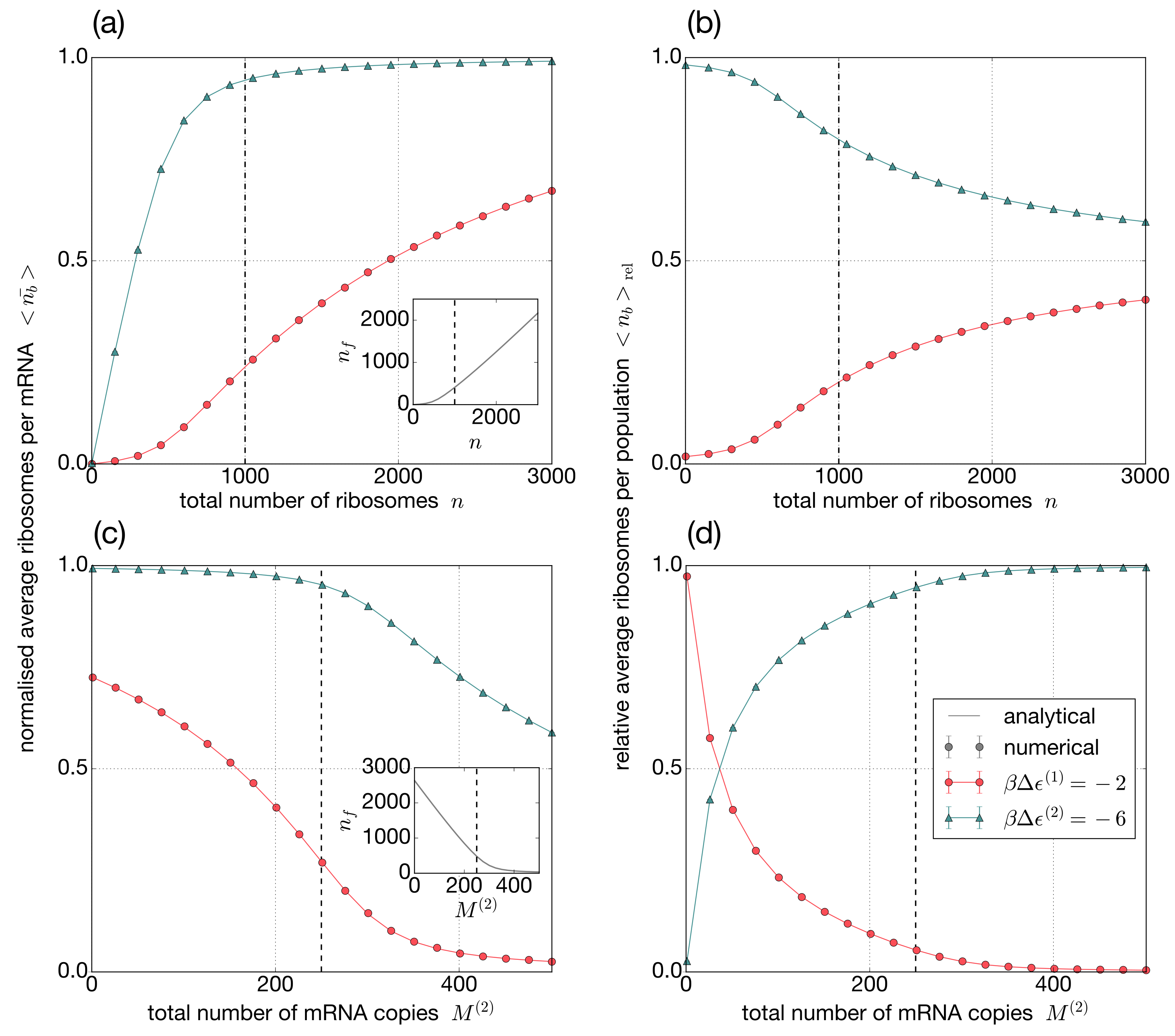}
		\caption{Two competing populations. Panels (a) and (b) respectively show $\big<\bar{n_b}\big>$ and $\big<n_b^{(i)}\big>_\mathrm{rel}$ as a function of $n$, for $N=10^4,\,\,n_{max}^{(i)}=10,\,\,M^{(i)}=50$. In panels (c) and (d) we vary the amount of transcripts $M^{(2)}$ of the population $2$. Dashed lines indicate when $\sum_{i}n_{max}^{(i)}M^{(i)}=n$, i.e. when the system enters a severely limited resources regime. Insets in (b) and (d) show the number of free ribosomes as functions of $n$ and $M^{(2)}$ respectively.} 
		\label{Case3_Fig1}
	\end{figure}
Instead of varying the supply of ribosome we can also change the amount of transcripts in each population. In panels (c) and (d) we increase $M^{(2)}$ and observe how the efficiency of each single transcript drops when entering a regime in which competition for ribosome is limiting (i.e. when $n<\sum_i  n_{max}^{(i)}M^{(i)}$), while the weight $\big<n_b^{(i)}\big>_\mathrm{rel}$ of the population $2$ increases at the expenses of the other population.

\section{Competition regulated gene expression}
Assuming that the translation rate of a transcript is proportional to the average number of ribosomes $\langle n_b^{(i)} \rangle$ bound to the population $i$, the evolution of the protein concentration $P^{(i)}$ produced by the population $i$ is given by
\begin{equation}
    \frac{d P^{(i)}}{dt} = k_\mathrm{TL}\langle n_b^{(i)} \rangle - \delta P^{(i)} \,,
\end{equation}
where $\delta$ is the degradation rate of that protein, and $k_\mathrm{TL}$ is the translation rate fixing the timescale of protein production of each ribosome. This gives (at steady-state) a protein concentration proportional to the total ribosomes engaged in the production of the protein of interest: $P^{(i)} = \frac{k_\mathrm{TL}}{\delta} \langle n_b^{(i)} \rangle$. \\

With these prescriptions and with the theory developed in the previous sections we can now speculate how competition for ribosomes can in principle affect cellular physiology and protein synthesis.

\subsection{Relative regulation of gene expression at the level of translation by ribosome competition}
We show that this model predicts that relative protein levels can be regulated by varying the supply or the demand of the translation machinery. Rather than absolute protein concentrations $P^{(i)}$ we also focus on relative expression levels $P^{(i)}_\mathrm{rel}$ defined as 
\begin{equation}
	P^{(i)}_\mathrm{rel} = \frac{P^{(i)}}{\sum_j P^{(j)}} = \langle n_b^{(i)}\rangle _\mathrm{rel} \,,
\end{equation}
and we assume that the translation and degradation rate of all proteins are the same. Thus, the relative protein abundance is equivalent to the fraction of ribosomes that are bound to the population of interest $i$, as defined in Eq.~(\ref{eq:rel_nb}). 

As a proof of principle, we plot the protein abundances of a system with $\mathcal{N}=3$ populations. This is similar to what we have shown in Fig.~\ref{Case3_Fig1}, but now we present a more complex situation with more populations competing for the same reservoir of ribosomes. We show that protein synthesis levels change when changing the trade-off of ribosomal resources. When $n \ll \sum_i n_{max}^{(i)}M^{(i)}$ the supply is smaller than the capacity of the system, and we are in a regime that is severely limited by the amount of ribosomes. In this regime all the transcripts are coupled via the pool of free ribosomes $n_f$. Otherwise, when $n \gg \sum_i n_{max}^{(i)}M^{(i)}$ the number of ribosomes is not limiting and asymptotically the system would behave as if each element is independent. 

	\begin{figure}[hbt] 
		\centering
		\includegraphics[width=1\textwidth]{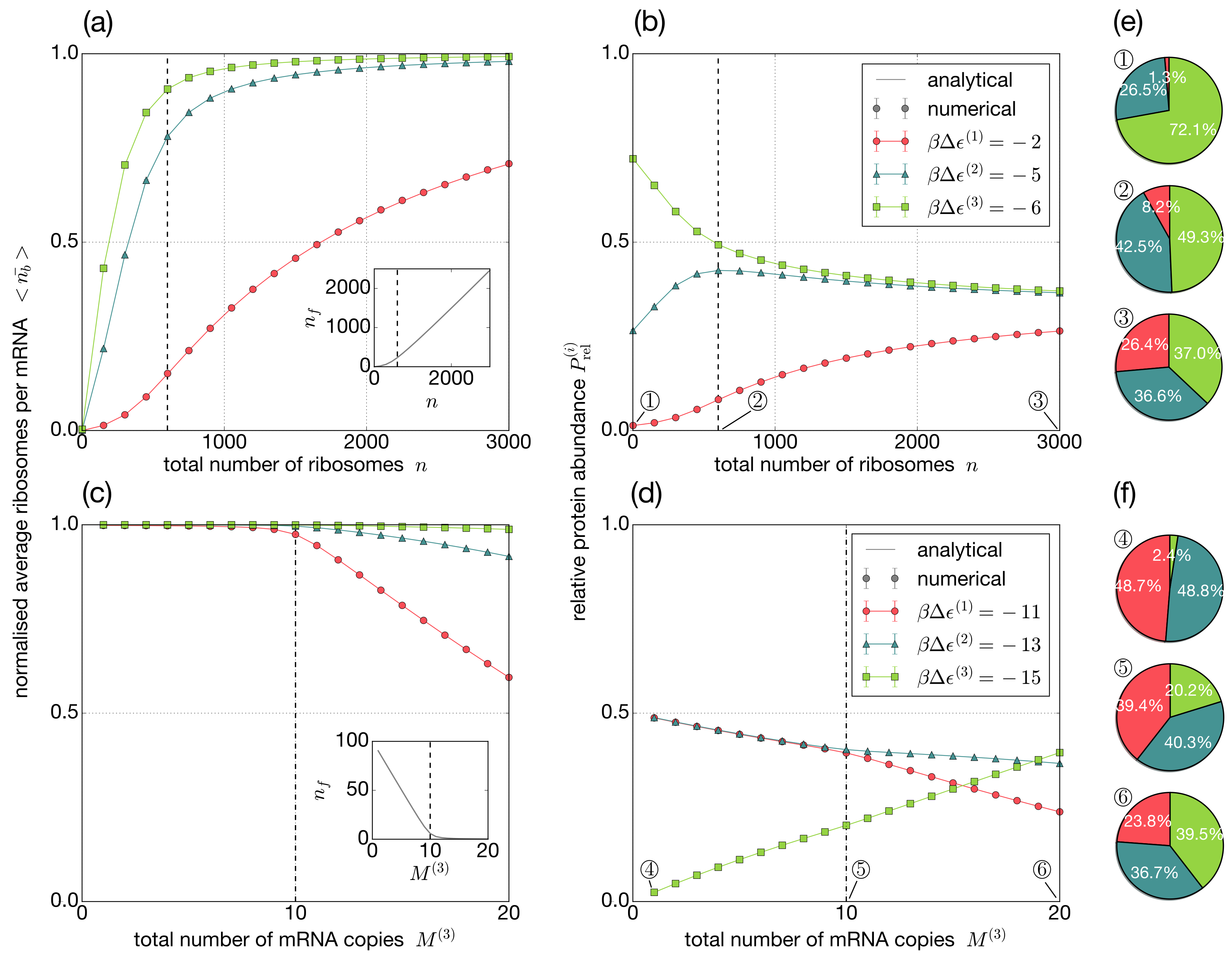}
		\caption{Finite ribosomal resources affect protein synthesis of competing populations ($\mathcal{N}=3$). In panel (a) we show the transcript efficiency $\big<\bar{n_b}\big>$ as a function of $n$, while in panel (b) the resulting relative composition of proteins. $M^{(i)}=20$ for each population and the energy turnover $\beta\Delta\epsilon$ is given in the legend of panel (b). Panels (c) and (d) show translation efficiency and the relative protein abundances when changing the amount of members of the population $3$ (energy turnover as in the legend of panel (d)).  $M^{(1)}= M^{(2)} = 20$. For all panels $N=10^4$, $n_\mathrm{max}^{(i)}=10$. Panels (e) and (f) show the pie charts of the protein composition for the points indicated in (b) and (d) respectively. } 
		\label{fig:regulation}
	\end{figure}

In panels (a) and (b) of Fig.~\ref{fig:regulation} we increase the supply of ribosomal resources $n$. We first plot the normalised ribosome occupation of each population, which is a proxy for the efficiency of protein production of each single mRNA. The efficiency of each transcript increases non-linearly until saturating for very large $n$, i.e. when resources are no longer limiting. However, mRNAs belonging to different populations increase their efficiency differently as a function of the supply available. To highlight this, in panel (b) we show the relative expression levels $P^{(i)}_\mathrm{rel}$, which is the fraction of proteins produced by the population $i$, as a function of the supply level. As the total number of ribosomes is increased, the relative expression changes in a counter-intuitive fashion. In the supply limited regime $n \ll \sum_i n_{max}^{(i)}M^{(i)}$, the strongest population (the one with the best transcript-ribosome affinity) engages most of the ribosomes, but when $n$ is increased its weight decreases at the advantage of the other populations. We stress that $P^{(i)}_\mathrm{rel}$ can present on optimal value or, in other words, that the production of given mRNA population can be optimised by tuning the amount of available number of ribosomes -see for instance the middle curve in panel (b) that shows a maximum.

In panels (c) and (d) of Fig.~\ref{fig:regulation} we instead vary the demand of the system by increasingly adding transcripts to the population $3$. When the system is in the non-limiting regimes (small $M^{(3)}$), the extra transcripts do not affect the efficiency of the mRNAs belonging to the other populations -panel (c)-, but the relative weights of their populations are already affected -panel (d)-. In the ribosome-limited regime the competition becomes severe and it drastically affects the efficiency of the individual mRNAs. Under strong competition the populations behave differently according to their recruitment capability.

We also represent in panels (e) and (f) the pie charts of the protein composition of this three-population system for three values of $n$ -panel (e)- and $M^{(3)}$ -panel (f).

\subsection{Relation between ribosome competition and cellular physiology}
In the previous section we have emphasised how ribosome competition might give rise to unexplored means of gene expression regulation at the level of translation. Now we speculate on the interplay between ribosome allocation between mRNAs of different populations and cellular physiology.

Following the derivation of~\cite{Scott2014}, the growth rate $\lambda$ of a cell is related to the biomass $\mathcal{M}$ and the total of actively translating ribosomes $n_\mathrm{act}$ by 
\begin{equation}
    \frac{d \mathcal{M}}{dt} = \lambda \mathcal{M} = k_\mathrm{TL} n_\mathrm{act} \,.
\end{equation}
Therefore, the growth rate is proportional to the number of ribosomes engaged in translation. Coarse graining the entire translatome of the cell with a single population of mRNAs, we can then identify $n_\mathrm{act}$ with $\langle n_b^{(1)}\rangle$, and thus assume that $\lambda \propto \langle n_b^{(1)}\rangle$. Figure~\ref{fig:physiology}(a) shows the relation between the total number of ribosomes and the growth rate, as predicted by our toy model. This almost linear regime, obtained with parameters that are physiologically reasonable, is qualitatively similar to the phenomenological growth law presented in~\cite{Scott2010}. Although we notice that $\lambda=0$ is obtained when $n=0$, one could impose a number of ribosomes $n_\mathrm{min}$ that cannot be involved in translation, $n_\mathrm{act} = \langle n_b^{(1)}\rangle - n_\mathrm{min}$, as done in~\cite{Scott2014}. 
	\begin{figure}[hbt] %
		\centering
		\includegraphics[width=1\textwidth]{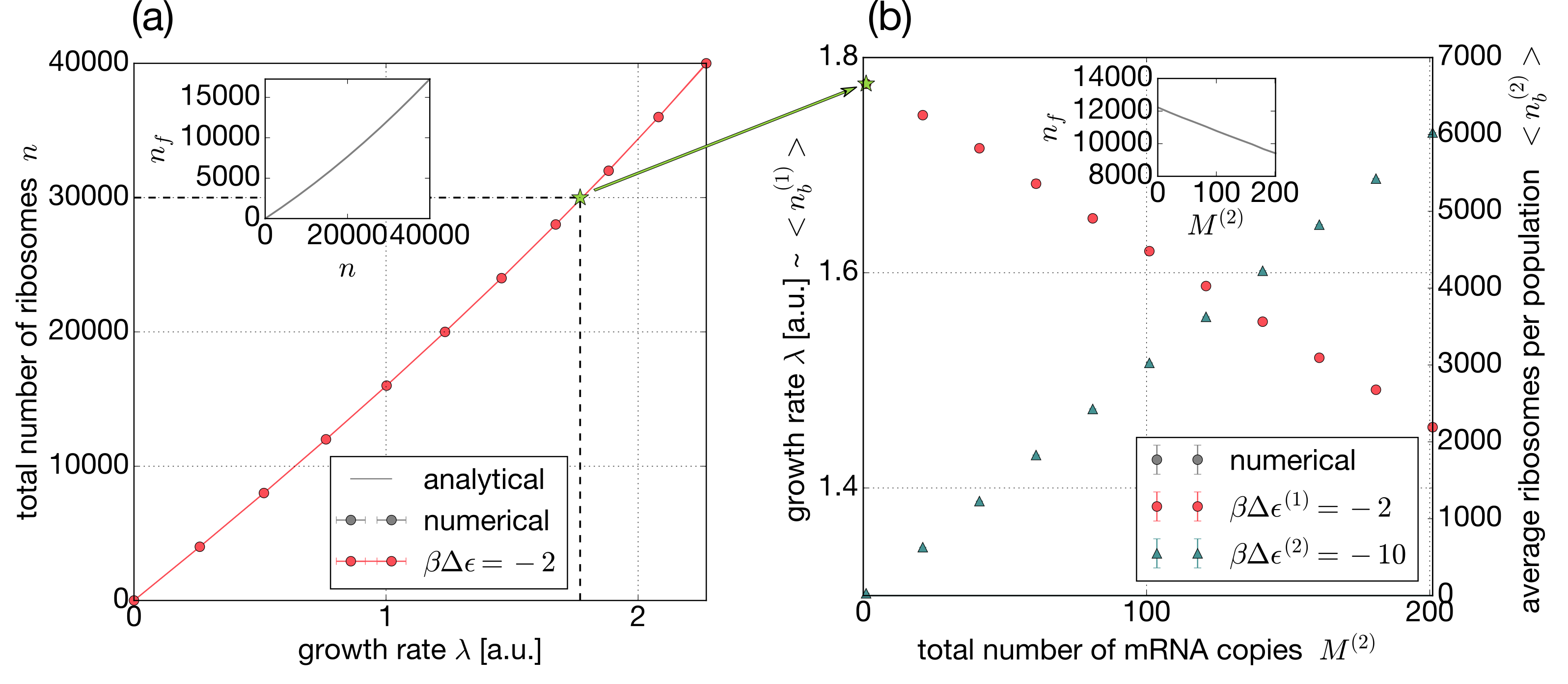}
 		\caption{Impact of ribosome competition on the cellular physiology. The relation between the estimated cellular growth rate and the total number of ribosomes is shown in panel (a) assuming a translatome composed of an individual population with $M^{(1)}=2000$, and energy turnover as in the legend. The inset shows the amount of free ribosomes $n_f$ as a function of $n$. By fixing the total amount of ribosomes to $n=3\times 10^4$ (green star), we imagine to start adding transcripts of an exogenous gene belonging to a second population, panel (b). The $M^{(2)}$ transcripts then increasingly soak up resources (i.e. the number of ribosomes $n_b^{(2)}$ engaged in translation of the exogenous population increases, triangles) that are necessary to the endogenous population, thus decreasing the growth rate (which is proportional to $n_b^{(1)}$, red circles). $M^{(1)}=2000$ as in panel (a). }
		\label{fig:physiology}
	\end{figure}
		
For a fixed number of ribosomes $n$, and thus for a given growth rate $\lambda$, the model predicts that when adding transcripts of an exogenous population $M^{(2)}$, the production of their protein will be at the expenses of the endogenous population, as shown in Fig.~\ref{fig:physiology}(b). Transcripts that are normally engaged in the translation of necessary endogenous proteins compete with the exogenous population for the pool of available ribosomes. The growth rate will then decrease as the demand of the competing population increases, and thus as the synthesis of the exogenous population grows. The mechanism proposed might be one of the basic principles of protein burden.

\section{Discussion}
In this work we have introduced an equilibrium model of mRNA translation to investigate, on theoretical grounds, the impact of ribosome competition on the translation machinery. We have extended the thermodynamic framework of gene expression, so far implemented to model transcription, to the translation step. 

Our toy model assumes that the ribosomal pool is in equilibrium with the translatome, the ensemble of actively translating mRNAs, and we allow for multiple ribosome binding on the same mRNA. Our coarse-grained approach neglects many biological processes, such as ribosome biogenesis or cellular metabolism. However, even such a simple model presents a complex phenomenology, and we speculate on qualitative aspects of ribosome competition related to gene expression regulation that are yet unexplored.

In principle the theory could be exploited for any number of mRNA populations, each one having different properties. However, the system can easily become intractable numerically because of the multiplicity terms $D_0(\{n_b^{(i)}\})$. Although the development of an approximated theory is out of the scope of this work, it is always possible to explore any possible system by means of the stochastic simulations explained in~\ref{appendix:Gillespie}.

This modelling framework could in principle be exploited in other processes, for instance to study the role of RNAP competition in transcription. Moreover, we can envisage further extensions of the model, for instance by considering the impact of temporal fluctuations in the supply-demand balance, or of cell-to-cell variability of ribosomes. The former could be investigated by including bursty transcription (for example by following the methodology explained in~\cite{Bokes2015ProteinSites, Soltani2015NonspecificProteins}), while we show numerical results of the latter in~\ref{asd}.

In this work we couple growth laws and gene expression via a mechanism of resource competition (for ribosomes). Previous coarse-grained models, see for instance~\cite{Lin2018HomeostasisCells}, assume that transcription and translation rates are proportional to relative abundances of genes and transcripts respectively, times the number of free RNAPs and ribosomes. Trade-off between these players is pivotal in determining the homeostasis of protein and mRNA concentrations in the cell during growth, and it explains different states of cell volume growth (exponential and linear). Instead, in this work we provide an exact formulation of the competition between mRNAs for ribosomes (or promoters for RNAPs), and obtain regimes of competition in which either ribosomes or transcript abundance are limiting. At the moment our analysis stops at the qualitative reproduction of the phenomenological growth laws, and a fine understating of the physiology at the single cell level is out of the scope of this work. However, finding more analytically tractable approximations of our model and applying frameworks similar to~\cite{Lin2018HomeostasisCells} would constitute an appropriate extension of our work.

We hope that our conjectural conclusions on relative gene expression regulation by finite resources and the interplay with cellular physiology may motivate experiments validating our results. Resource competition has been so far largely overlooked in the literature, and we expect that such theoretical works might encourage practical applications in synthetic biology and even in {\it in vitro} systems, where supply and demand can be better controlled and decoupled from the cellular physiology.

\section{Acknowledgements}
LC and PSR would like to thank the Erasmus exchange program, thanks to which PSR could spend time at the University of Montpellier working on this project, and Prof. Joerg Fitter who was PSR's contact at the origin institute. TJR acknowledges Fondecyt Iniciacion 11161046 for funding.

%
%
\section*{References}
\bibliographystyle{unsrt}
\bibliography{pascal}

\appendix

\section{Supporting the analytical solutions using simulations based on the Gillespie algorithm\label{appendix:Gillespie}}

We implement a simulation scheme based on the `direct method' version of Gillespie algorithm  ~\cite{Gillespie1977ExactReactions, Rydenfelt2014StatisticalTitration} to support the analytical results obtained in Section \ref{subs:cases} and computed by numerically solving the equations throughout the text.
By implementing this chemical reaction algorithm that connects reacting species (ribosomes and mRNAs) with the corresponding chemical reactions (un/binding of ribosomes and mRNAs), we obtain a stochastic simulation in excellent agreement with the rather complicated and detailed analytical solutions. 
The Gillespie algorithm draws the time $\tau$ of the next possible reaction from an exponential distribution $e^{-A\tau}$ with mean equal to the inverse of the sum of all the reaction rates of possible events $A=\sum_{j}a_j$, with $a_j$ being the rate of the event $j$ - also named propensity. The subsequent reaction $r$ is then again randomly drawn from the uniform distribution with probability density function $1/A$. \\

In order to derive the necessary kinetic rates we first focus on the case $\mathcal{N} = M = n_{max} = 1$. The reaction between the unbound (with $n$ free ribosomes) and bound (with $n-1$ free ribosomes) state can be written as  
\begin{equation}
n \,; n_b = 0  \underset{k_{off}}{\stackrel{k_{on}}{\rightleftharpoons}} n-1\,; n_{b}=1
\end{equation}
where $k_{on}$ and $k_{off}$ are reaction rate constants. Hereby, $k_{on}$ is the rate of number of associations per free ribosomes, per mRNA, and $k_{off}$ is the rate of disassociation of the bound particle, per free compartments. We find these constants by solving the chemical master equation of the probability of occupation $p(n_b=1)$: 
\begin{equation}
\frac{d}{d t} p(n_b=1) =  k_{on} p(n_b=0) n - k_{off} p(n_b=1) (N - n + 1)\,\,. 
\end{equation}
At the steady state, with the occupation probability $p$ according to the case described in Section \ref{subsubs:cases_singlecopy} we obtain the relation
\begin{equation}
k_{on} = k_{off}\frac{(N-n+1)}{n} \frac{p(n_b=1)}{p(n_b=0)}=k_{off} e^{-\beta\,\Delta\epsilon}\,\,. \label{appendix:gillespie2}
\end{equation}

The exponential weight accounts for the RBS strength, meaning that different populations $i$ of mRNAs react on different time scales due to specific $k_{on}^{(i)}(\Delta\epsilon^{(i)})$. \\
After having computed the kinetic constants $k_{on}^{(i)}$ and $k_{off}^{(i)}$ for each population $i$, and since those constants only depend on the property of the mRNA, the Gillespie scheme for each mRNA $\alpha$ of the population $i$ can be written as
\begin{equation}
n_f \,; n_b^{(i,\alpha)} \underset{k_{off}^{(i)}}{\stackrel{k_{on}^{(i)}}{\rightleftharpoons}} n_f -1 \,; n_{b}^{(i,\alpha)}+1 \,\,,
\end{equation}
with $n_b^{(i,\alpha)} < n_{max}^{(i)}$. In the previous equation we have introduced the notation with the index $(i,\alpha)$ to highlight the binding/unbinding event for each mRNA $\alpha$, and we emphasise that the rate constants are the same for all the mRNAs in the population $i$. Thus the individual propensities are $a_{on}^{(i,\alpha)} = n_f \, k_{on}^{(i)}(n_{max}^{(i)}-n_b^{(i,\alpha)})$ and $a_{off}^{(i,\alpha)}=k_{off}^{(i)}n_b^{(i,\alpha)} (N-n_f)$. 
For the sake of simplicity one could also write the algorithm in terms of the population $i$, without distinguishing the individual mRNAs. In this case the propensities for each population are $a_{on}^{(i)} = n_f \, k_{on}^{(i)}(n_{max}^{(i)} M^{(i)}-n_b^{(i)})$ and $a_{off}^{(i)}=k_{off}^{(i)}n_b^{(i)} (N-n_f)$.

The algorithm follows the time evolution of the chemical species $n_b^{(i,\alpha)}$. After a transient time, the abundances of species tend to equilibrium. We then calculate the time average over several intervals of the same length and compute the average value and its standard deviation. We choose $10^7$ iteration steps to overcome the transient region, and calculate the average value with 10 intervals of $10^6$ iterations. \\ \\

%
\section{Determination of $N$
\label{appendix:determinationN}}
For comparison with a physiological relevant system, we suggest an approximate value of the number of boxes $N$ for the bacterial \textit{E.coli}. In Section \ref{subs:thermodynamic} we mention the cells' cytoplasm for distributing the free ribosomes but did not go into a more detailed description, which we provide here. \\
In order to compute the number of available boxes we consider that the cytoplasmic volume $V_{cyto}$ contributes to the total number of boxes. Reasonably assuming ribosomes as spheres, the effective stacking of these (see Figure \ref{fig:sketch} (a)) leads to about $30\%$ of space not accessible for distribution, by which we correct to $0.7\,V_{cyto}$ of available space for the spheres. We further approximate that $V_{cyto}\approx V_{cell}$, where $V_{cell}$ is the volume of the cell, as the width of the membrane is negligible in comparison with the `radius' of the cell. Hence $N\approx 0.7\,\,(V_{cell}/V_{ribo})$, where $V_{ribo}$ is the volume of a ribosome. \\
The cell volume $V_{cell}\approx 1.1\mu $m$^3$ corresponds to $n\approx 3\cdot10^4$ ribosomes~\cite{Milo2016CellNumbers}. With the ribosome volume $V_{ribo}\approx 3.4\cdot10^{-6} \mu $m$^3$~\cite{Verschoor1998Three-dimensionalRibosome}, we then find $N\approx 2.3\cdot10^5$, whereby about $20\%$ is already occupied by the $n$ ribosomes. \\ \\
%
\section{Derivation of the mathematical results as given in the general case
\label{appendix:DerivationGeneralCase}}
In Section \ref{subs:thermodynamic} we introduced the partial partition function $Z$. Here we derive the normalized partial partition function $Z_0$ for the general case with $\mathcal{N}$ populations each one composed of $M^{(i)}$ transcripts. $Z=Z(\{n_b^{(i)}\})$ is a function of the bound particles $n_b^{(i)}$ to all populations $i$ and determined by the sum of all microstates (the degeneracy $D$) with the same exponential weight (according to the systems energy state $E$):
\begin{equation}
Z(\{n_b^{(i)}\})= D(\{n_b^{(i)}\}) \, e^{-\beta E(\{n_b^{(i)}\})} \,.
\label{appendix:degeneracy_Eq1}
\end{equation}
One can find the energy of a state with $\{n_b^{(i)}\}$ by summing up the energetic contributions of bound and unbound particles, with $\epsilon_b$ and $\epsilon_f$ respectively, to be $E(\{n_b^{(i)}\})=[n_f\epsilon_f+\sum_{i=1}^{\mathcal{N}}n_b^{(i)}\epsilon_b^{(i)}]$. Using $n_f=n-n_b=n-\sum_{i}n_b^{(i)}$ one finds $E(\{n_b^{(i)}\})=[n\epsilon_f+\sum_{i=1}^{\mathcal{N}}n_b^{(i)}(\epsilon_b^{(i)}-\epsilon_f)]=[n\epsilon_f+\sum_{i=1}^{\mathcal{N}}n_b^{(i)}\Delta\epsilon^{(i)}]$, with the change in energy $\Delta\epsilon^{(i)}$ given by the chemical reaction between a ribosome and an mRNA of population $i$. \\
The contributions to the non-normalized degeneracy $D(\{n_b^{(i)}\})$ are, as described in Section~\ref{subs:thermodynamic}, on the one hand the degeneracy of distribution $n_f=n-n_b$ of free particles into the cytoplasm with $N$ boxes, and on the other hand the product of the degeneracy of $n_b^{(i)}$ bound ribosomes on the $n_{max}^{(i)}M^{(i)}$ sites of population $i$. This gives
\begin{equation}
	D(\{n_b^{(i)}\}) = {N \choose n-n_b} \prod_{i=1}^{\mathcal{N}}{n_{max}^{(i)}M^{(i)} \choose n_b^{(i)}}\,\,.
	\label{appendix:degeneracy_Eq2}
\end{equation}
As mentioned above, the degeneracy of the populations are decoupled (leading to the product over the populations), as a change of particle between populations leads to a change of $E$, representing different macrostates. \\
Plugging $D$ and $E$ in Equation (\ref{appendix:degeneracy_Eq1}) we obtain the partial partition function of the general case of $i=1,...,\mathcal{N}$ different populations with $M^{(i)}$ copies 
\begin{equation}
Z(\{n_b^{(i)}\})= {N \choose n-n_b} \prod_{i=1}^{\mathcal{N}}{n_{max}^{(i)}M^{(i)} \choose n_b^{(i)}} \, e^{-\beta [n\epsilon_f+\sum_{i=1}^{\mathcal{N}}n_b^{(i)}\Delta\epsilon^{(i)}]}\,\,.
\label{appendix:degeneracy_Eq3}
\end{equation}
To normalise Equation (\ref{appendix:degeneracy_Eq3}) it seems reasonable to choose as reference state $n_b=0$ for which $\{n_b^{(i)}\}=\{0\}$. With the corresponding 
\begin{equation}
Z(\{0\})= {N \choose n}  \, e^{-\beta [n\epsilon_f]}\,\,.
\label{appendix:degeneracy_Eq4}
\end{equation}
one obtains
\begin{eqnarray}
Z_0(\{n_b^{(i)}\})&=\frac{Z(\{n_b^{(i)}\})}{Z(\{0\})}& \\
&= \frac{{N \choose n-n_b}}{{N \choose n}} \prod_{i=1}^{\mathcal{N}}{n_{max}^{(i)}M^{(i)} \choose n_b^{(i)}} \, e^{-\beta [\sum_{i=1}^{\mathcal{N}}n_b^{(i)}\Delta\epsilon^{(i)}]}& \\
&=D_0(\{n_b^{(i)}\}) \, e^{-\beta [\sum_{i=1}^{\mathcal{N}}n_b^{(i)}\Delta\epsilon^{(i)}]}&\,\,,
\label{appendix:degeneracy_Eq5}
\end{eqnarray}
where $D_0$ is the degeneracy refered to state $n_b=0$. The quantity $D_0$ has further been reduced and expressed as given in Equations (\ref{Case3_Eq3}) and (\ref{Case3_Eq4}) of Section \ref{subsubs:cases_generalcase} to make it more accessible for computational purposes. For doing so, one applies the following relations
\begin{eqnarray*}
\hspace{-12ex}
(1)\qquad\qquad& \frac{b!}{(b-c)!}=b\cdot(b-1)\cdot...\cdot(b-c+1)=
&\left\{
\begin{array}{ll}
\prod_{j=1}^{c}(b-j+1) & c >0\\
1 & c = 0 
\end{array} 
\right.\\
\hspace{-12ex}(2)&\frac{(a-b)!}{(a-(b-c))!}=\frac{1}{(a-b+1)\cdot...\cdot(a-b+c)}=
&\left\{
\begin{array}{ll}
\prod_{j=1}^{c}\frac{1}{a-b+j} & c >0\\
1 & c = 0
\end{array}
\right.\\
\hspace{-12ex}(3)&\frac{1}{c!}=\frac{1}{1\cdot 2\cdot...\cdot c}=
&\left\{
\begin{array}{ll}
\prod_{j=1}^{c}\frac{1}{j} & c >0\\
1 & c = 0
\end{array}
\right.\,\,.
\end{eqnarray*}

\section{Cell-to-cell variability of ribosomal resources}
\label{asd}
Although in this work we focused on fixed values of resources (constant number of ribosomes), in this Appendix we perform a simulation by considering a population of isogenic cells whose ribosome content is Gaussian distributed (and mRNA is constant). Figure~\ref{fig:colony} shows the outcome of this simulation, which corresponds to the situation shown in Figure~\ref{fig:regulation} with normally distributed total amount of ribosomes $n$ (standard deviation 5\%).
\begin{figure}[h!bt]
	\centering
	\includegraphics[width=1\textwidth]{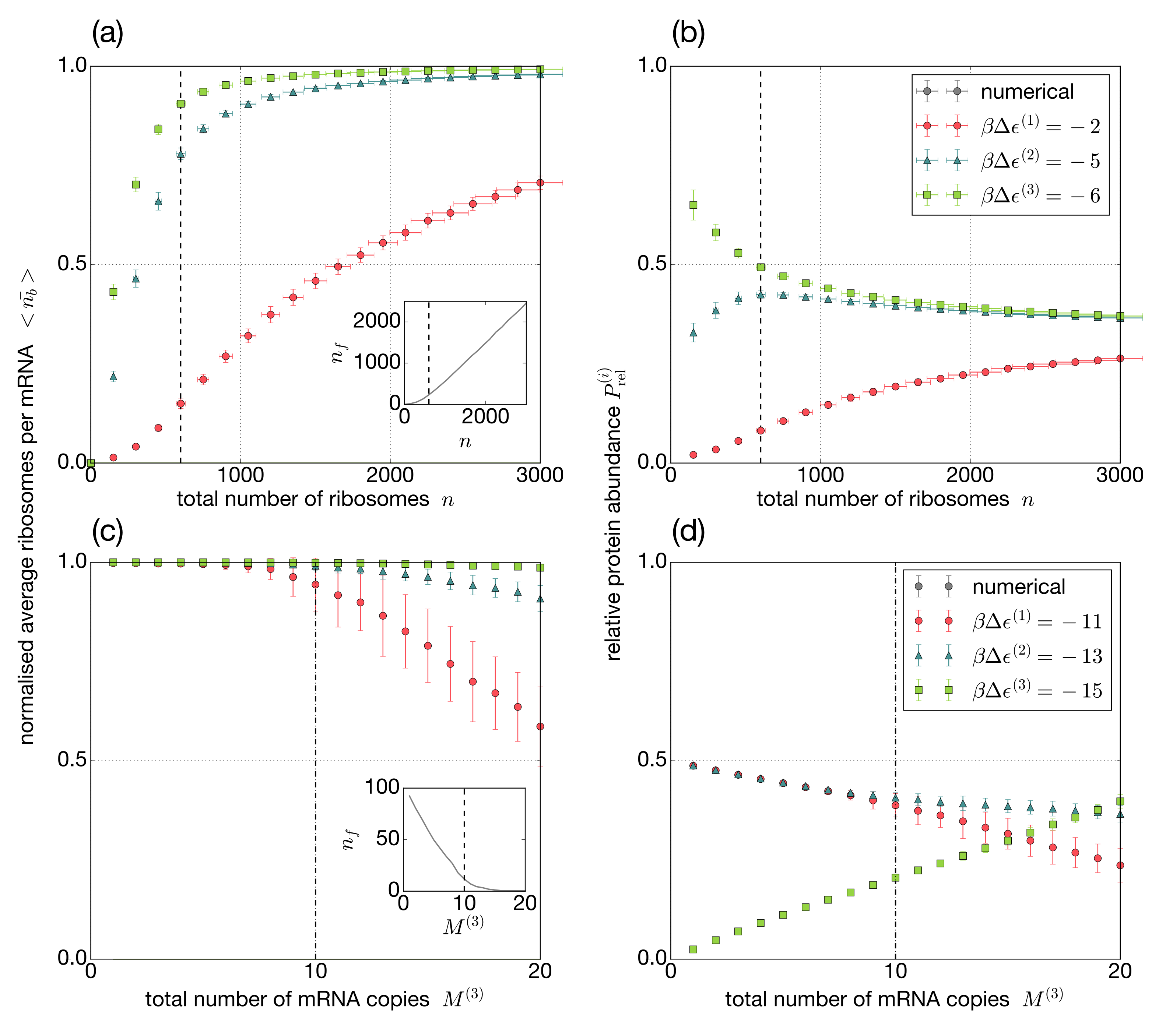}
		\caption{Equivalent of Figure~\ref{fig:regulation} but with normally distributed $n$, standard deviation 5\%. Each point is the result of 100 simulations with $n$ drawn from this distribution.}
    \label{fig:colony}
\end{figure}

\section{Alternative mRNA degeneracy}
In the main text we have coarse-grained each mRNA population and characterised it as a 2D-lattice -see Fig \ref{fig:sketch} (b)- with $n_{max}^{(i)}M^{(i)}$ accessible sites. With this choice we cannot follow each transcript individually, and the macrostate of a population $i$ with $n_b^{(i)}$ ribosomes bound has a degeneracy ${n_{max}^{(i)}M^{(i)}}\choose{n_b^{(i)}}$.

However, other representations of ribosome recruitment and of the mRNA population are possible. For instance, we could refine the treatment of the mRNA populations by considering each transcript independently, which is equivalent to alternatively represent the degeneracy terms $D_0^*$.  

In order to extend the model and explicitly consider each transcript in a population, we recall a well known problem in combinatorics: ``We consider two dice each having six different numbers of eyes. What is the number of different ways to roll the two dice, given that we are interested in the sum of the eyes? Note that the order of rolling the dice is not of concern for our considerations.'' This problem can be translated into the language of our problem: The sum of the eyes is the quantity $n_b^{(i)}$, the number of dice is $M^{(i)}$ and the number of different numbers of eyes is $n_{max}^{(i)}+1$. In the formulation of the problem used in the text we substantially defined the problem with a single dice per population. Instead, the solution of this combinatorial problem has been found by Abraham de Moivre a few centuries ago, and we obtain:

\begin{equation}
D_0^*(i) = \sum_{k=0}^{\lfloor{n_b^{(i)}/(n_{max}^{(i)}+1)}\rfloor}(-1)^k\, {M^{(i)} \choose k}\, {{n_b^{(i)}+M^{(i)}-k\,(n_{max}^{(i)}+1)-1}\choose{M^{(i)}-1}}\,\,.
\label{appendix:alternativeDeg}
\end{equation}

We have tested this approach in a small system composed of $\mathcal{N} = 3$ populations each one composed of $M^{(i)} = 5$ mRNAs as shown in Figure~\ref{fig:extension}. In panel (a) we show the relative ribosomes per mRNA population $\langle n_b \rangle_\mathrm{rel}$ as computed with the model definitions used in the main text. Instead, in panel (b) we compute the same quantity with the degeneracy introduced in this appendix. When the system is in a regime not limited by the resources (large $n$), the two models converge to the same result. However, there are evident differences when ribosomes are limiting. If further extensions are out of the scope of this paper, here we show that small changes in the model can largely impact the outcome. This could be probably used to understand what are the main relevant rules that should be included in the model when comparing the theory to experimental results. 

Finally, we remark that the de Moivre's solution matches with the Gillespie simulations only for relatively small systems. We observed a deviation between theory and simulations for larger systems, that we believe is mainly due to numerical error propagation.

\begin{figure}[h!bt]
	\centering
	\includegraphics[width=1\textwidth]{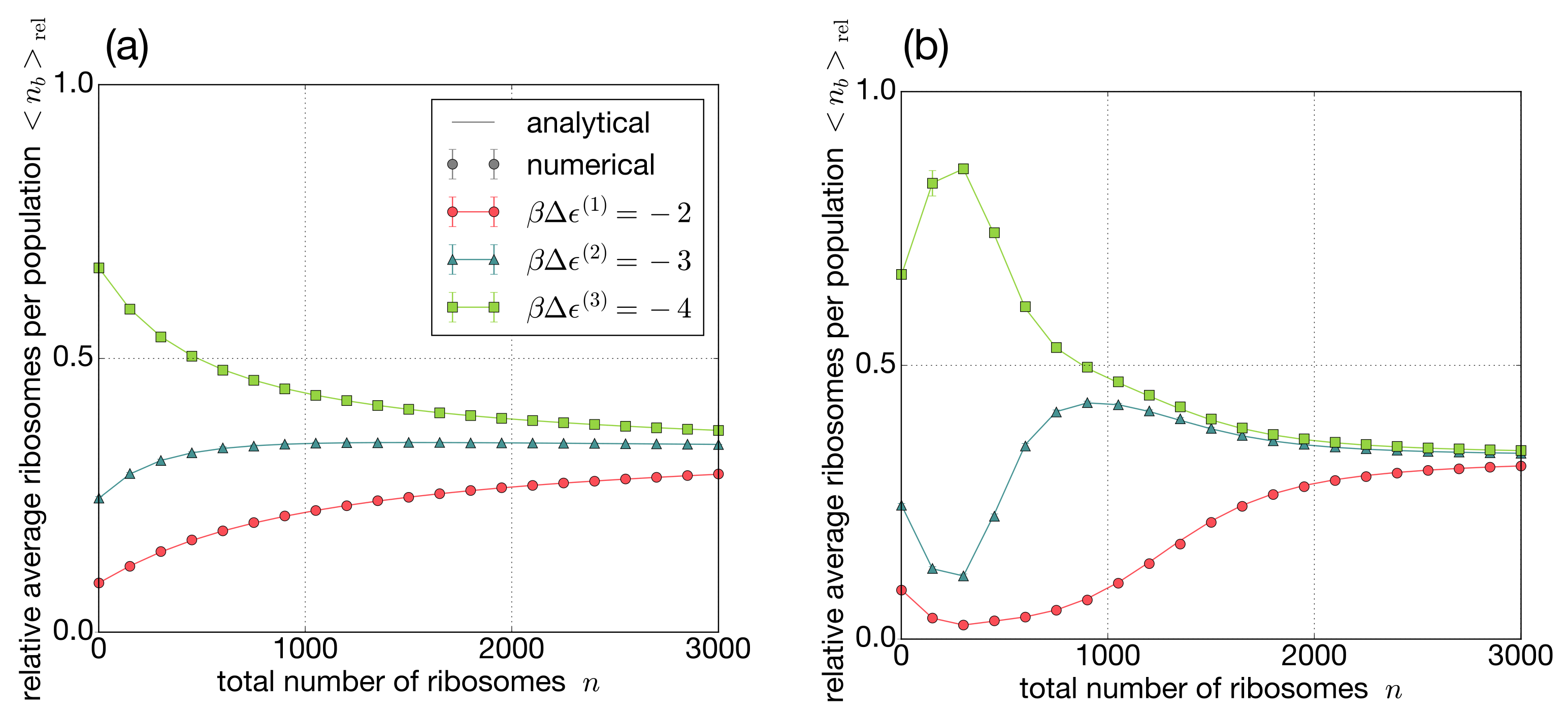}
		\caption{Relative ribosomes per population with the model definitions used in the main text -panel (a)- and with the prescriptions given in this appendix -panel (b)-. $N=10^4$, $n_{max} = 10$. }
    \label{fig:extension}
\end{figure}

\end{document}